# Nonvolatile magneto-thermal switching driven by vortex trapping in commercial In-Sn solder

Poonam Rani[1], Takumi Murakami[1], Yuto Watanabe[1], Hossein Sepehri-Amin[2], Hiroto Arima[1,3], Aichi Yamashita[1], Yoshikazu Mizuguchi[1]*

1. Department of Physics, Tokyo Metropolitan University, 1-1, Minami-osawa, Hachioji 192-0397, Japan.
2. National Institute for Materials Science, 1-2-1, Sengen, Tsukuba 305-0047, Japan.
3. National Institute of Advanced Industrial Science and Technology, 1-1-1 Umezono, Tsukuba 305-8563, Japan.

## ABSTRACT

Magneto-thermal switching (MTS) is a key technology for efficient thermal management. Recently, large MTS with nonvolatility has been observed in Sn-Pb solders [H. Arima et al., Commun. Mater. 5, 34 (2024)] where phase separation, different superconducting transition temperature ($T_c$) of Sn and Pb, and magnetic-flux trapping are the causes of the nonvolatile MTS. To further understand the mechanism and to obtain the strategy for enhancing switching ratio, exploration of new phase-separated superconductors with nonvolatile MTS is needed. Here, we show that the In52-Sn48 commercial solder is a phase-separated superconducting composite with two $T_c$ and traps vortices after field cooling. A clear signature of nonvolatile MTS was observed at $T$ = 2.5 K. From specific heat analyses, we conclude that the vortices are mainly trapped in the lower-$T_c$ phase (γ-phase) after field cooling, which is the evidence that vortex trapping also works on achieving nonvolatile MTS in phase-separated superconducting composites.

**Introduction**

Superconductivity is a quantum phenomenon that emerges at low temperatures. At temperatures ($T$) lower than the superconducting transition temperature ($T_c$) electrons form Cooper pairs, and electrical resistivity becomes zero in the superconducting states [1]. In addition, in the superconducting states, electronic thermal conductivity ($\kappa$) is suppressed because the Cooper pairs do not transfer heat [2]. The reduced $\kappa$ in the superconducting states can be recovered to high $\kappa$ by the application of magnetic field ($H$) greater than critical field ($H_c$) or upper critical field ($H_{c2}$) of the material. Using the change in $\kappa$ by $H$, magneto-thermal switching (MTS) can be achieved. MTS is a key technology in the field of thermal management [3,4] because the MTS can achieve heat flow control without any mechanical motions [5–7]. Recently, we reported that large MTS ratio can be achieved in pure-element superconductors (Nb and Pb) with high purity [8–10]. Although the working temperature should be low because of low $T_c$, the MTS using low-$T_c$ superconductors will be useful for thermal management of low-$T$ devices [11,12], if the performance can be further improved. Furthermore, to make the MTS useful, it is necessary to develop materials with nonvolatile MTS in which high (or low) $\kappa$ is retained even after removing external $H$. In Nb with intermediate state, nonvolatility was observed at low $T$ [2], and the origin was explained as the affection of magnetic flux to phonon scattering. In the case of Nb intermediate state, the $H$ experience resulted in lower $\kappa$. Recently, we achieved controllable large nonvolatile MTS in Sn-Pb solders [13]. In the Sn-Pb solder, the nonvolatile MTS is driven by the change in electronic $\kappa$ by the external $H$, and the switching ratio and the absolute value of $\kappa$ can be controlled by Sn-Pb concentration and $H$ [13,14]. The Sn-Pb solders are known as phase-separated composites composed of Sn and Pb; both are type-I superconductors with $T_c$ = 3.7 and 7.2 K and $H_c$ (0 K) ~ 300 and 800 Oe, respectively [15,16]. By field-cooling (FC) or reducing $H$ from $H > H_c$, magnetic fluxes are trapped in the Sn regions, and bulk superconductivity of Sn is suppressed because of the trapped field greater than $H_c$ of Sn [13]. Similarly, magnetic flux trapping in MgB$_2$ ($T_c$ = 39 K) causes nonvolatile MTS [17], but the switching ratio is smaller than that of Sn-Pb solders. Therefore, other examples of nonvolatile MTS using a superconducting transition has been desired to be studied because enrichment of examples of nonvolatile MTS will provide us with strategies for achieving a higher switching ratio of nonvolatile MTS. Here, we show the results on characterization, physical properties, and nonvolatile MTS characteristics of commercial Sn52-In48 solder, which is used in low-$T$ soldering. The superconducting properties of In-Sn solders have been reported in Ref. 18, and the $T_c$ of the solder is higher than that of pure In ($T_c$ = 3.4 K) and Sn. According to the binary In-Sn phase diagram, phase separation into a β-phase (In-rich: $T_c$ = 6.5 K) and γ-phase (Sn-rich: $T_c$ = 4.7 K) [18]. Because of alloying states of those phases, the emergence of type-II superconductivity, in which vortices are formed, in both regions are expected [19–22]. If the vortex states can be retained after FC and demagnetization, and if nonvolatile MTS emerges, the material design space for nonvolatile MTS materials will be expanded. Therefore, we performed detailed characterization on the In52-Sn48 solder to obtain new insights on magnetic flux trapping and nonvolatile MTS in superconductor composites.

**Results**

Fig. 1 shows the X-ray diffraction (XRD) pattern of the In52-Sn48 plate made by pressing the solder sample. A clear phase separation into the β-phase (In-rich, $I4/mmm$, No. 139) and γ-phase (Sn-rich, $P6/mmm$, No. 191) is seen, and no other impurity phases were observed. The crystal structure of the β-phase is same as pure In. Figure 2 shows the elemental mapping results based on scanning-electron microscope (SEM) and energy-dispersive X-ray spectroscopy (EDX). As shown in Figs. 2c and 2d, clear and homogeneous phase separation into the two phases was observed. From the composition line analysis, the atomic composition of the β-phase can be estimated as $In_{0.68(2)}Sn_{0.32(2)}$, and that for the γ-phase is $In_{0.30(2)}Sn_{0.70(2)}$, which is consistent with the known phase diagram. The noticeable feature of the phase separation of the In52-Sn48 solder is presence of square-shaped γ-phase regions surrounded by the β-phase regions. The shape is different from the Sn-Pb solders studied in Ref. 13. However, the situation of lower-$T_c$ regions surrounded by higher-$T_c$ regions is similar to that of the Sn-Pb solders.

Figure 3a shows the $T$ dependence of magnetization ($4\pi M$) after zero-field cooling (ZFC) and FC with an applied magnetic field of $H$ = 10 Oe. Large diamagnetic signals are observed at $T$ < 6.5 K, which is consistent with the $T_c$ of the β-phase. There is no double-step transition, while the In52-Sn48 sample contains two superconducting samples. This trend is quite similar to that observed in Sn-Pb solders, and the absence of double-step transition would be explained by the micro-scale phase separation and proximity effects [13]. Figure 3b shows the $T$ dependence of $4\pi M$ measured at $H$ = 0 Oe after FC at various $H$. By FC at 500 Oe, at $T$ = 1.8 K and at $H$ = 0 Oe, magnetic field of about 450 Oe is trapped. Furthermore, by FC at $H \geq$ 1000 Oe, the trapped field of 620 Oe was observed, which indicates that the flux trapping saturates at $H$ = 1000 Oe. Figures 4c and 4d show the $H$ dependence of $4\pi M$ and inner magnetic field ($B$) at $T$ = 2.5 K.

To investigate superconducting properties, specific heat was measured at different field conditions. Figure 4a shows the $T$ dependence of specific heat ($C$) in the form of $C/T$ measured at $H$ = 0 Oe after ZFC and FC (3000 Oe). The $C$ measurements were performed three times at each $T$. For the FC data, first data point is masked because of anomalous sample heating related to flux reduction (anomalous temperature rise in the $C$ measurements), which is the trend similar to Sn-Pb solders [14]. There is slight difference in $C/T$ between ZFC and FC data. To highlight the difference, normal-state $C$ ($C_n$), which is shown in Fig. 4b, is subtracted from the $C/T$ data. Figure 4c shows the $T$ dependence of $(C-C_n)/T$ calculated using $C$ measured at $H = 0$ Oe after ZFC and FC (3000 Oe). At around $T = 6$ K, the superconducting signal is observed for both data, which indicate that the superconducting states of the ZFC and FC states are comparable down to 4.2 K. Therefore, the FC states of the β-phase would be close to Meissner states above $T = 4.2$ K. Below 4.2 K, there is a clear difference between the ZFC and FC data. In the ZFC data, there is a clear and sharp superconducting transition possibly of the γ-phase. In contrast, for the FC data, the superconducting transition of the γ-phase is clearly suppressed; the $T_c$ is lower and the entropy change is smaller than that observed in the ZFC data. These results imply that the superconducting states of the γ-phase are affected by the presenting magnetic field, which is protected by the supercurrents of the β-phase, but the γ-phase is still bulk superconducting. The superconducting states of the γ-phase should be type-II state with vortices, and this trend is the clear difference from the case of Sn-Pb solders. The schematic image of the presence of vortices in the square-shaped γ-phase regions is displayed in Fig. 4d.

As characterized in the above result part, the presence of type-II bulk superconducting states with vortices in the γ-phase surrounded by the β-phase with the Meissner state is clearly different from the situation of Sn-Pb solders where Sn is not bulk superconducting (when evaluating from $C$ measurement). Therefore, if we could observe nonvolatile MTS in the In52-Sn48 solder sample, the material design space will be largely expanded because of the availability of type-II superconductors as a component. Figure 5a shows the $T$ dependence of $κ$ measured at $H = 0$ (superconducting state) and 3000 Oe (normal-conducting state). There is a clear difference, and MTS can occur in the In52-Sn48 solder sample. In Fig. 5b, the $H$ dependence of $κ$ measured at $T = 2.5$ K after initial ZFC is shown with a clear nonvolatility. The difference in $κ$ between the before and after field experience (nonvolatile MTS ratio) is 45%.

**Discussion**

Here, we investigated structural, compositional, and physical properties of the In52-Sn48 commercial solder. The solder traps a large amount of flux as observed in the Sn-Pb solders, but both phases are bulk superconducting even after FC, which indicates that, at least, the γ-phase is in type-II superconducting states with vortices. We observed nonvolatile MTS in the solder, and the switching ratio was 45%, which is smaller than typical values observed in the Sn-Pb solders (150% for Sn45-Pb55 and 300% for Sn10-Pb90). However, we notice that the $\kappa$ of the In52-Sn48 commercial solder is quite low, which would be the reason why the nonvolatile MTS ratio was limited. Therefore, further optimization of In-Sn composition and material production processes would improve the MTS characteristics of In-Sn solders. At the end, but not least, the observation of nonvolatile MTS in a phase-separated superconducting composite containing a type-II superconductor expands the material-exploration space because of the availability of various superconductors. That is quite important for developing superconducting nonvolatile MTS using high-$T_c$ superconductors.

**Methods**

The investigated In52-Sn48 (mass ratio In:Sn = 52:48) commercial solder wires (Chip Quik Inc.) with a diameter of 0.79 mm. XRD was performed on a pressed sample in a plate form using a RIGAKU diffractometer Miniflex-600 with a Cu-Kα radiation by the $\theta$-$2\theta$ method. SEM and EDX were performed using Carl Zeiss Cross-Beam 1540ESB for the micro-structure analysis with elemental mapping on the surface of the solder. Thermal conductivity ($\kappa$) was measured using Physical Property Measurement System (PPMS, Quantum Design) with a thermal transport option (TTO) using a four-probe steady-state method with a heater, two thermometers, and base-temperature terminal. The lengths between two thermometers attached to the measured samples was 25.1 mm. Due to the limitation of the sample-room space of the TTO stage, the sample was screwed to store inside with four probes, a heater, two thermometers, and thermal base. The typical measurement duration for a single measurement was 30 seconds. Magnetization was measured using superconducting quantum interference device (SQUID) magnetometry on Magnetic Property Measurement System (MPMS3, Quantum Design) with a VSM mode. Specific heat was measured on PPMS by a relaxation mode. The sample was attached on a stage using APIEZON N grease.

**Figures**

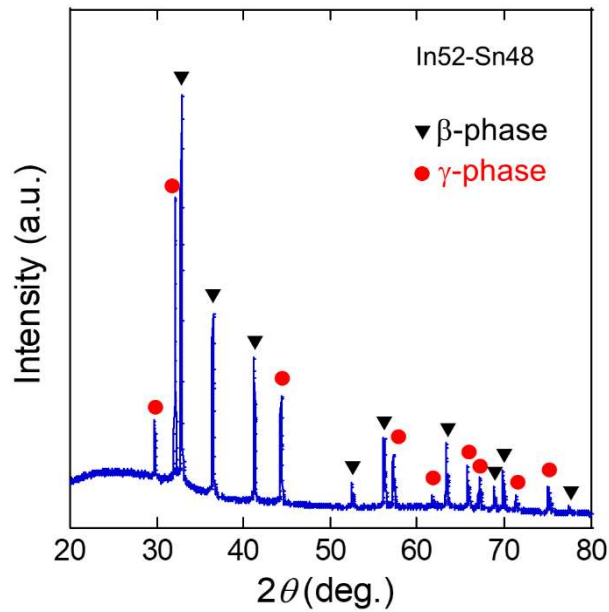

**Fig. 1 | X-ray diffraction pattern of In52-Sn48.** The filled red circles and black triangles indicate the peaks for the γ-phase and β-phase, respectively.

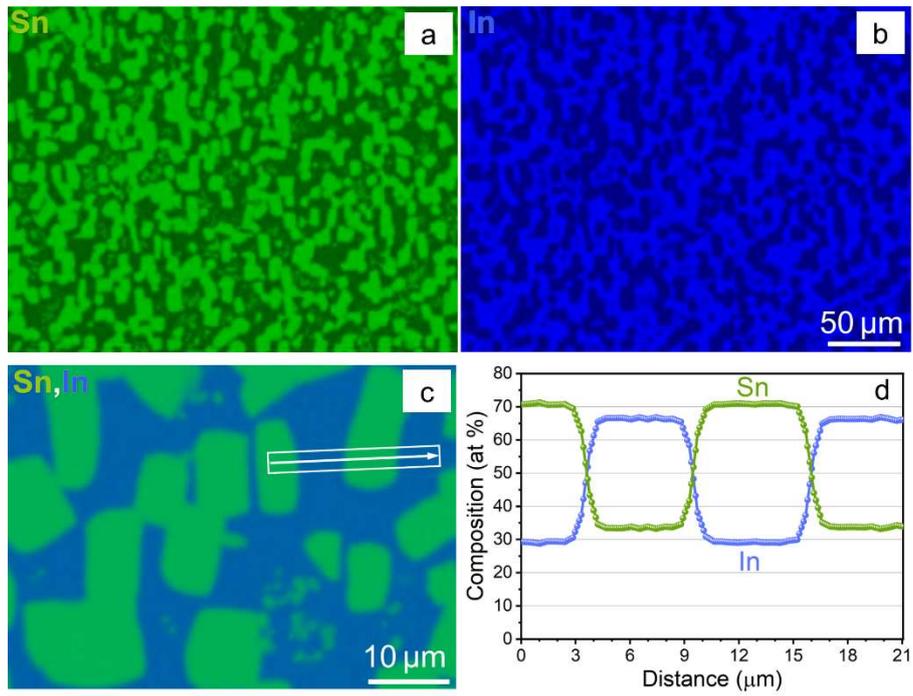

**Fig. 2 | Elemental mapping. a,b** SEM-EDX elemental mapping for Sn and In. **c** Sn/In contrast mapping. **d** Composition line profile of constituent elements performed along the arrow in (c).

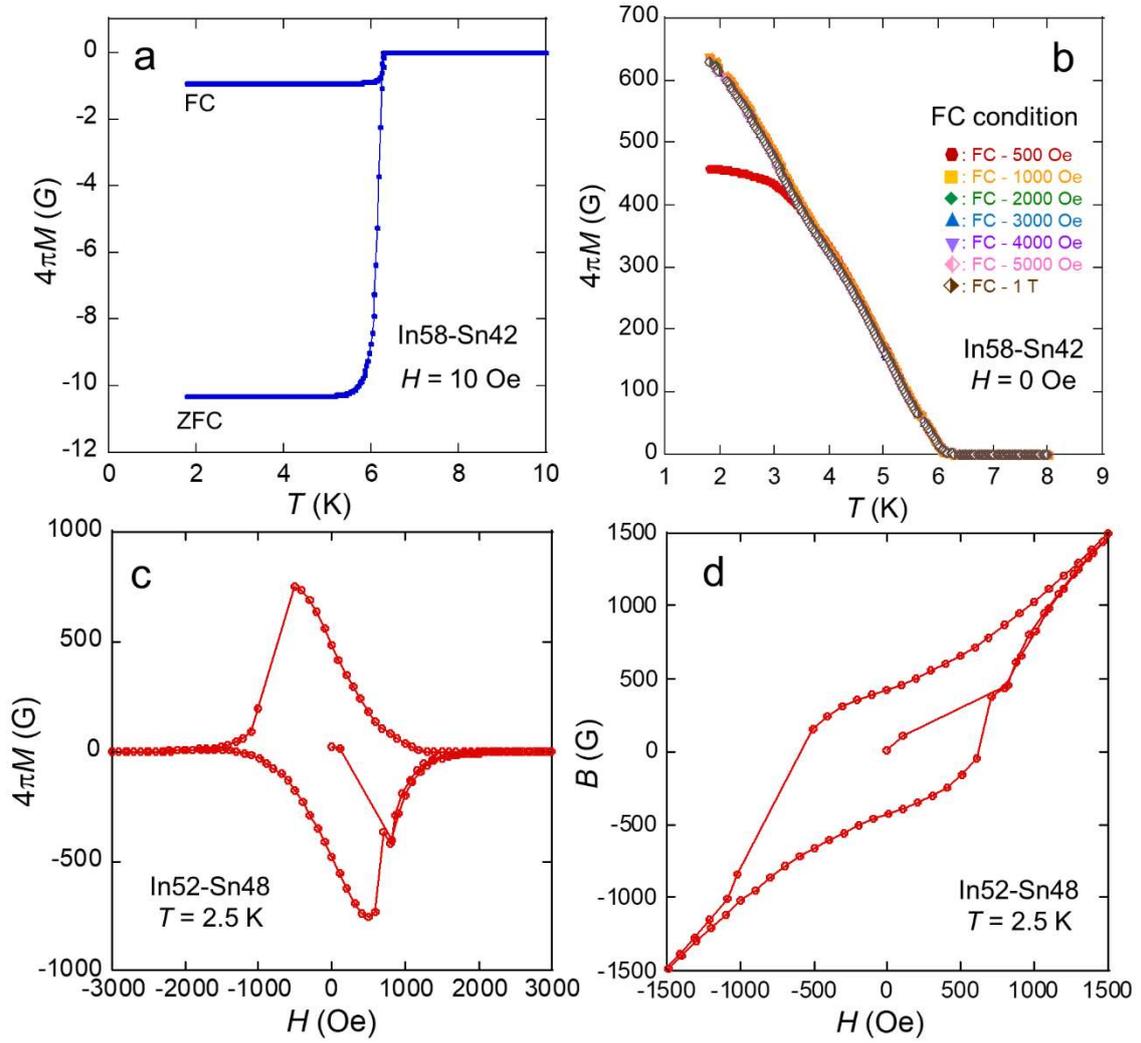

**Fig. 3 | Magnetic properties of In52-Sn48. a** $T$ dependence of $4\pi M$ after ZFC and FC with an applied field of $H$ = 10 Oe. **b** $T$ dependence of $4\pi M$ measured at $H$ = 0 Oe after FC at various $H$. **c,d** $H$ dependence of $4\pi M$ and inner magnetic field ($B$) at $T$ = 2.5 K.

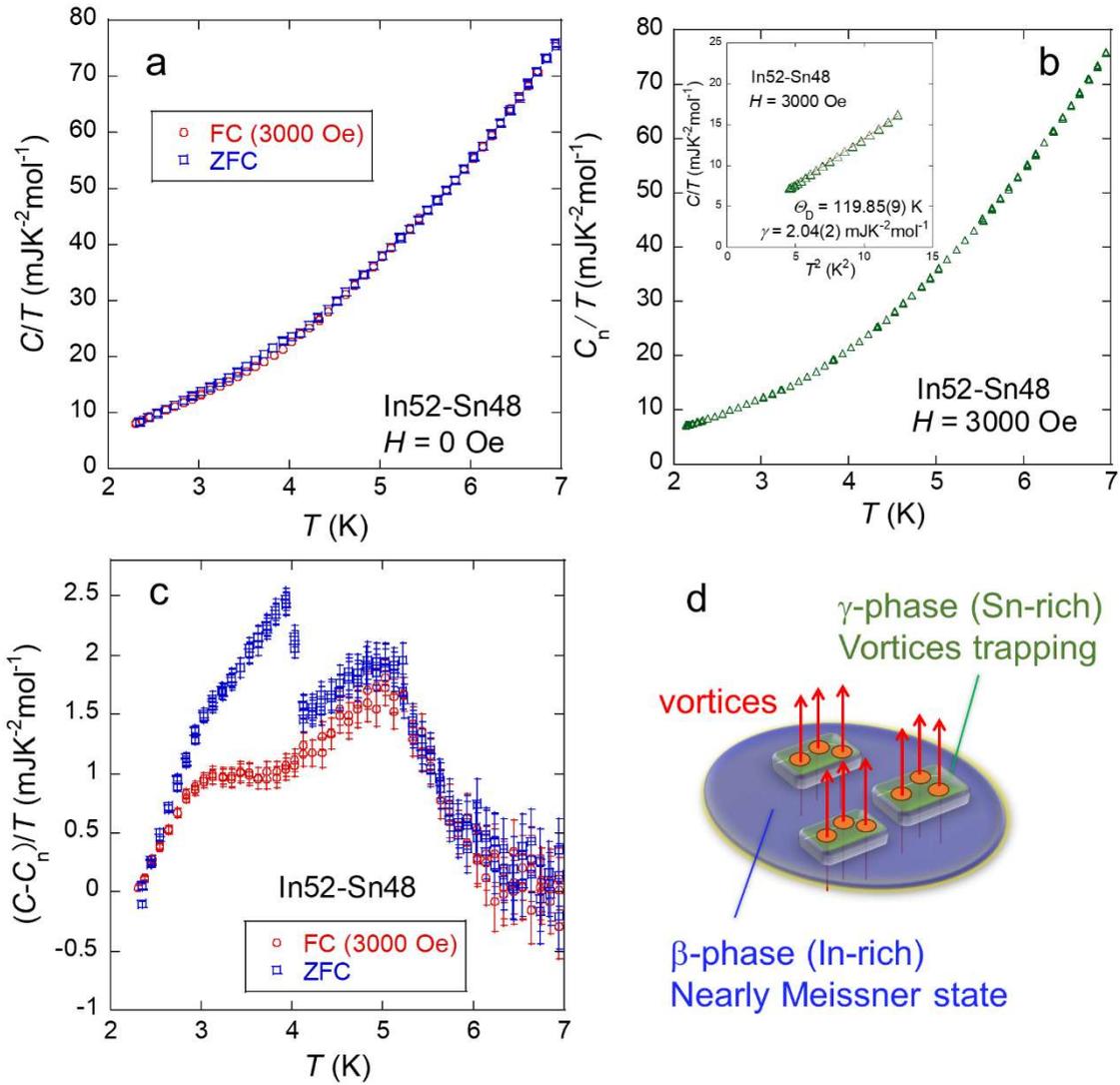

**Fig. 4 | Specific heat data for In52-Sn48. a** $T$ dependence of $C/T$ measured at $H = 0$ Oe after ZFC and FC (3000 Oe). **b** $T$ dependence of normal-state specific heat $C_n/T$. **c** $T$ dependence of $(C-C_n)/T$ calculated using $C$ measured at $H = 0$ Oe after ZFC and FC (3000 Oe). **d** Schematic image of vortex trapping at the γ-phase regions.

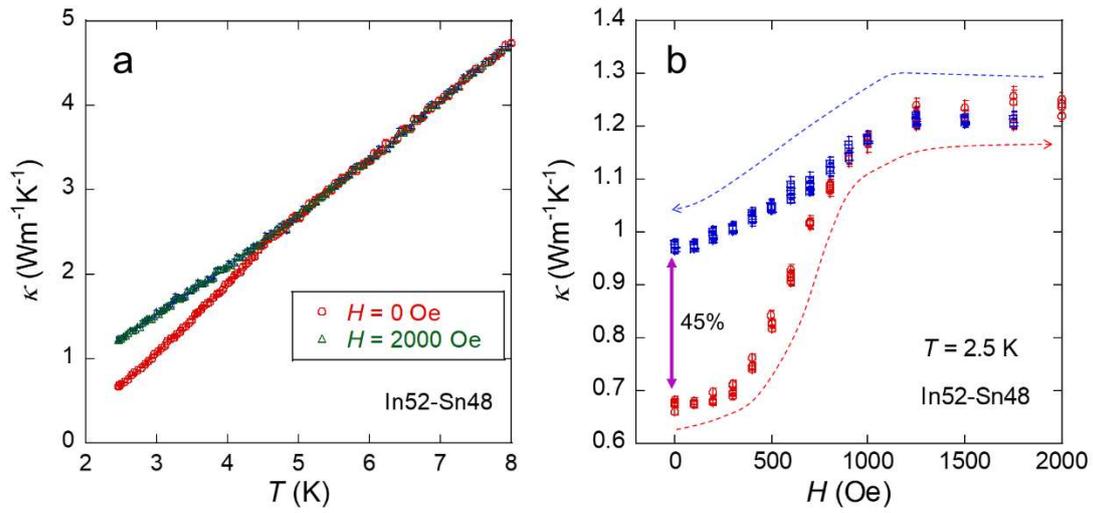

**Fig. 5 | Nonvolatile MTS of In52-Sn48. a** $T$ dependence of $\kappa$ measured at $H = 0$ and 3000 Oe. **b** $H$ dependence of $\kappa$ measured at $T = 2.5$ K after ZFC.

**Acknowledgements**

The authors thank F. Ando, K. Uchida, T. Ichikawa for discussion on the results. This project was partly supported by TMU Research Project for Emergent Future Society and JST-ERATO (JPMJER2201).